\documentclass[preprint,notoc]{JHEP3}
\usepackage{epsfig}

\def\eslt{\not\!\!{E_T}}
\def\eslt{E_T^{\rm miss}}

\def\to{\rightarrow}

\def\bi{\begin{itemize}}
 \def\ei{\end{itemize}}
\def\te{\tilde e}
\def\c1p{C1^\prime}
\def\ta{\tilde a}
\def\tG{\tilde G}

\def\tu{\tilde u}

\def\ta{\tilde a}

\def\tb{\tilde b}

\def\tst{\tilde t}

\def\tg{\tilde g}

\def\tq{\tilde q}

\def\tw{\tilde\chi}

\def\twpm{\tilde\chi^\pm}
\def\tz{\tilde\chi^0}
\def\alt{\stackrel{<}{\sim}}
\def\agt{\stackrel{>}{\sim}}
\def\be{\begin{equation}}  
\def\ee{\end{equation}}  
\def\bea{\begin{eqnarray}}  
\def\eea{\end{eqnarray}}  

\title{Prospects for Yukawa Unified $SO(10)$ SUSY GUTs \\ 
at the CERN LHC}
\author{Howard Baer$^{a}$, Sabine Kraml$^b$, Sezen Sekmen$^c$ 
and Heaya Summy$^{a,d}$\\
$^a$Dept.\ of Physics and Astronomy, University of Oklahoma, Norman, OK 73019, USA\\
$^b$Laboratoire de Physique Subatomique et de Cosmologie, UJF Grenoble 1, 
CNRS/IN2P3, INPG, 53 Avenue des Martyrs, F-38026 Grenoble, France\\
$^c$Department of Physics, Middle East Technical Univ., TR-06531 Ankara, Turkey\\
$^d$Department of Physics, Florida State University, Tallahassee, FL 32306, USA\\
E-mail: \email{baer@nhn.ou.edu}, \email{sabine.kraml@lpsc.in2p3.fr},
\email{sezen.sekmen@cern.ch}, \email{heaya@hep.fsu.edu}}

\preprint{\vbox{FSU-HEP-080822, LPSC 08-133}}

\abstract{
The requirement of $t-b-\tau$ Yukawa coupling unification
is common in simple grand unified models based on the gauge group
$SO(10)$, and it also places a severe constraint on the expected 
spectrum of superpartners.
For Yukawa-unified models with $\mu >0$, the spectrum is characterized by 
three mass scales:
{\it i}).~first and second generation scalars in the multi-TeV range, 
{\it ii}).~third generation scalars, $\mu$ and $m_A$ in the few-TeV range and 
{\it iii}).~gluinos in the $\sim 350-500$ GeV range with 
chargino masses around $100-160$ GeV. 
In such a scenario, gluino pair production should occur at large rates at the CERN LHC, 
followed by gluino three-body decays into neutralinos or charginos.  
Discovery of Yukawa-unified SUSY at the LHC 
should hence be possible with only 1 fb$^{-1}$ of integrated 
luminosity, by tagging multi-jet events with 2--3 isolated leptons,  
without relying on missing $E_T$. 
A characteristic dilepton mass edge should easily be apparent above 
Standard Model background. Combining dileptons with $b$-jets, along with
the gluino pair production cross section information, should  allow
for gluino and neutralino mass reconstruction. A secondary corroborative 
signal should be visible at higher integrated luminosity in the
$\twpm_1\tz_2\to 3\ell$ channel, and should exhibit the same
dilepton mass edge as in the gluino cascade decay signal.
}

\begin{document}

\section{Introduction}
\label{sec:intro}

Models for new physics based on $SO(10)$ grand unification and supersymmetry
are especially compelling\cite{review}. $SO(10)$ grand unified theories (GUTs) allow for--
in addition to gauge coupling unification--
{\it matter unification} in the sense that all fields of a single generation
of the Standard Model (SM) are embedded into the 16-dimensional spinor
representation of $SO(10)$. Since there are only 15 fields in a SM generation, 
the 16th element is occupied by a right-handed neutrino (RHN) field.
The RHN is necessary for implementing the see-saw mechanism\cite{seesaw} of generating 
neutrino masses. In addition, since the gauge group $SO(10)$ is anomaly free,
the ad-hoc but fortuitous triangle anomaly cancellation in the SM and in $SU(5)$ GUTs
is explained. The addition of supersymmetry to the model allows for a stabilization
of the weak scale-GUT scale hierarchy, and is supported by the fact that the measured
weak scale gauge couplings actually unify at $Q=M_{GUT}\simeq 2\times 10^{16}$ GeV
under minimal supersymmetric standard model (MSSM) renormalization group 
evolution (RGE). Along with gauge and matter unification, the simplest $SO(10)$ SUSY GUT models
may also have {\it Yukawa coupling unification}, especially for the third generation.

On the minus side, $SO(10)$ SUSY GUT models in 4 dimensions are often unwieldy, due in part to 
the large Higgs representations which are needed to break the GUT gauge symmetry.
In recent years, progress has been made in constructing SUSY GUT theories in
5 or 6 spacetime dimensions\cite{exdimguts}. 
In these cases, the GUT symmetry can be broken by extra-dimensional
compactification on for instance an orbifold. Such theories maintain many of the essential
predictions of SUSY GUTs, while dispensing with GUT-breaking via large Higgs representations.
In any case, if we consider that the MSSM arises from superstring theory as a 4-d
effective theory below $M_{GUT}$, then 6-7 extra dimensions will have to be compactified, 
and the description of string$\to$GUT$\to$MSSM is likely to be more complicated
than any of the existing models. 

To avoid dealing with the unknown physics above the GUT scale, 
in this paper we will assume that an $SO(10)$ SUSY GUT model breaks (either via the Higgs mechanism or via
compactification of extra dimensions) to the SM gauge group at energy 
scale $Q=M_{GUT}$. Below $M_{GUT}$, we assume the MSSM is the correct effective 
field theory which describes nature. We will further assume that the superpotential
above $M_{GUT}$ is of the form
\be
\hat{f}\ni f\hat{\psi}_{16}\hat{\psi}_{16}\hat{\phi}_{10} +\cdots
\ee 
so that the three third generation Yukawa couplings $f_t$, $f_b$ and $f_\tau$ are
{\it unified at $M_{GUT}$}.\footnote{It is simple in this context to include as well the effect of
a third generation neutrino Yukawa coupling $f_\nu$; this effect has been shown
to be small, although it can affect Yukawa coupling unification by a few per cent, if the right 
hand neutrino mass scale is within a few orders of magnitude of $M_{GUT}$\cite{abbbft}.}
We will further make the reasonable assumption that the GUT scale soft SUSY breaking terms
are constrained by the $SO(10)$ gauge symmetry such that matter scalars have a 
common mass $m_{16}$, Higgs scalars have a common mass $m_{10}$ and there is a common 
trilinear soft parameter $A_0$. 
The bilinear soft term $B$ can be traded as usual for $\tan\beta$, the ratio of Higgs 
field vevs, while the magnitude of the superpotential Higgs mass $\mu $ is 
determined in terms of $M_Z^2$ via the electroweak symmetry breaking 
minimization conditions. Here, electroweak symmetry is broken radiatively (REWSB) 
due to the large top quark mass.

In order to accomodate REWSB, it is well-known that the GUT scale Higgs soft masses must be
{\it split} such that $m_{H_u}^2 <m_{H_d}^2$, in order to fulfill the EWSB minimization conditions.
Two possibilities-- full $D$-term splitting applied to all scalar masses (DT), or splitting applied only to
Higgs soft terms (HS)-- have been considered\cite{bdr2,abbbft}. 
The latter HS case was found to yield a higher degree
of Yukawa coupling unification when $m_{16}\sim 10$ TeV, so that this is the case we will consider here.  
We parametrize the Higgs splitting as $m_{H_{u,d}}^2=m_{10}^2\mp 2M_D^2$. 
The Higgs splitting can originate via a large near-GUT-scale threshold correction arising from the
neutrino Yukawa coupling: see Ref. \cite{bdr2} for discussion. Thus, the Yukawa unified
SUSY model is determined by the parameter space
\be
m_{16},\ m_{10},\ M_D^2,\ m_{1/2},\ A_0,\ \tan\beta, \ sign(\mu )
\label{eq:space}
\ee
along with the top quark mass, which we take to be $m_t=171$ GeV.\footnote{Recent
measurements from CDF and D0\cite{mtop} now find a central value for $m_t$ of $172.6\pm 1.4$ GeV.}

In previous work, the above parameter space was scanned over (via random scans\cite{bf,abbbft} and also
by more efficient Markov Chain Monte Carlo (MCMC) scans\cite{bkss}) to search for Yukawa unified solutions.
The quantity
\be
R=\frac{max(f_t,f_b,f_\tau )}{min(f_t,f_b,f_\tau )},
\label{eq:R}
\ee
was examined, where solutions with $R\simeq 1$ gave apparent Yukawa coupling unification.
For superpotential Higgs mass parameter $\mu >0$ (as favored by $(g-2)_\mu$ measurements),
Yukawa unified solutions with $R\sim 1$ were found
but only for {\it special} choices of GUT scale boundary conditions\cite{bf,bdr1,bdr2,abbbft,drrr,bkss}:
\be
A_0\sim -2m_{16},\ m_{10}\sim 1.2 m_{16},
\label{eq:BCs}
\ee
with $m_{1/2}\ll m_{16}$ and $\tan\beta \sim 50$.
In fact, models with this sort of boundary conditions were derived even earlier in the context
of inverted scalar mass hierarchy models (IMH) which attempt to reconcile suppression of
flavor-changing and $CP$-violating processes with naturalness 
via multi-TeV first/second generation and sub-TeV scale third generation scalars\cite{bfpz}.
The Yukawa-unified spectral solutions were thus found in Refs. \cite{abbbft,bkss} to occur with the 
above peculiar choice of boundary conditions along with $m_{16}\sim 3-20$ TeV 
and $m_{1/2}\sim 50-100$ GeV. 
For accurate spectrum evaluation, we adopted the Isajet 7.75 algorithm\cite{isajet,bfkp}, which includes 
\bi
\item full two-loop RG evolution of both couplings and soft SUSY breaking (SSB) terms\cite{mv},
\item minimization of the RG-improved one-loop scalar potential at an optimized scale $M_{SUSY}$
(we take $M_{SUSY}=\sqrt{m_{\tst_L}m_{\tst_R}}$), which accounts for leading two-loop terms\cite{hhh},
\item implementation of complete 1-loop $t$, $b$ and $\tau$ threshold corrections\cite{hrs,bmpz} and
\item implementation of complete 1-loop corrections for all sparticle masses\cite{bmpz}.
\item a hybrid approach of changing gauge and Yukawa coupling beta functions as different soft terms
are integrated out of the effective theory\cite{bfkp}.
\ei
A plot of Yukawa coupling evolution for case A of Table 1 of Ref. \cite{bkss} with $m_{16}\simeq 9$ TeV 
is shown in Fig.~\ref{fig:yu}. Here, we clearly see the importance of the MSSM theshold corrections, 
which give rise to the discontinuities located around $Q=M_{SUSY}\sim 3$ TeV.
\FIGURE[htb]{
\epsfig{file=yuk.eps,width=10cm} 
\caption{ Plot of $f_t$, $f_b$ and $f_\tau$ 
evolution from the weak scale to the GUT scale
for Point A of Table \ref{tab:bm}. The large jumps around 3 TeV 
correspond to the MSSM threshold corrections.
}\label{fig:yu}}

Based on the above work\cite{abbbft,bkss}, the sparticle mass spectra from Yukawa-unified SUSY 
models are characterized by the following conditions:
\begin{itemize}
\item first and second generation scalars have masses in the $3-20$ TeV regime, 
\item third generation scalars, $\mu$ and $m_A$ have masses in the $1-4$ TeV regime (owing to the
inverted scalar mass hierarchy), 
\item gluinos $\tg$ have mass in the $350-500$ GeV range,
\item light charginos have mass $m_{\tw_1}\sim 100-160$ GeV and 
\item the lightest neutralino $\tz_1$ is nearly pure bino with mass $m_{\tz_1}\simeq 50-80$ GeV.
\end{itemize}

The presence of a bino-like $\tz_1$ along with multi-TeV scalars gives rise to a neutralino
dark matter (DM) relic abundance that is typically in the range $\Omega_{\tz_1}h^2\sim 10^2-10^5$, {\it i.e.}
far above\cite{auto} the WMAP measured\cite{wmap3} value $\Omega_{CDM}h^2=0.111^{+0.011}_{-0.015}$ ($2\sigma$) 
by several orders of magnitude.

A very compelling way out for solutions with $m_{16}\sim 5-20$ TeV, 
is to {\it not} assume that the lightest neutralino $\tz_1$ is the lightest SUSY particle (LSP). 
Instead, we assume that the 
{\it axino} $\ta$ (the spin-${1\over 2}$ element of the supermultiplet containing the axion)
is in fact the LSP\cite{covi}. 
The axino mass is very model-dependent, but can be anywhere in the keV-GeV range\cite{axmass,covi}.
In this case, the neutralino mass over-abundance is greatly reduced since 
$\tz_1\to \ta\gamma$ decay can occur with a lifetime of order $0.03$ sec. 
This decay time is sufficiently short that late-time neutralino decay to axino in the early universe
should not upset successful predictions of Big Bang Nucleosynthesis (BBN).
The relic abundance then gets reduced by the 
ratio $m_{\ta}/m_{\tz_1}$ which can be of order $10^{-2}-10^{-5}$. The $\ta$ coming from $\tz_1$ decay would actually
constitute {\it warm} dark matter\cite{jlm}. However, axinos can also be produced thermally in the early universe as 
cold dark matter\cite{covi,steffan}. 
Thus, in the axino LSP scenario, we would actually have dominantly thermally produced cold axino DM with a small
admixture of warm axino DM arising from $\tz_1\to\ta\gamma$ decays.
In addition, there is also the possibility of a significant presence of {\it axion} CDM.

In fact, in this class of Yukawa-unified solutions with $m_{16}$ in the multi-TeV range, we also expect the
gravitino $\tG$ to lie in the multi-TeV range. The cosmological gravitino problem-- wherein gravitinos produced
thermally in the early universe suffer a late-time decay, thus destroying the successful predictions
of BBN-- can be avoided. For $m_{\tG}\alt 5$ TeV, the re-heat temperature $T_R$
must be $T_R\alt 10^5$ GeV\cite{kohri}, thus creating tension with most viable mechanisms for baryogenesis\cite{buchm}. 
However, for
$m_{\tG}\agt 5$ TeV, the re-heat bound is much higher: $T_R\alt 10^8-10^9$ GeV. 
This range of $T_R$ is exactly what is needed for baryogenesis via {\it non-thermal} leptogenesis\cite{ntlepto},
wherein the heavy right-hand neutrino states are not produced thermally, but rather via inflaton decay.
It also turns out to be the exact range needed to generate a dominantly {\it cold} axino DM universe. Thus, the
whole scenario fits together to offer a consistent cosmological picture of BBN, baryogenesis and cold DM
composed of axinos\cite{bs}!

An alternative but equally compelling solution to the CDM problem in Yukawa-unified SUSY models 
occurs in cases where $m_{16}$ is as low as the $\sim 3$ TeV range.
In this case, the $\tz_1$ can remain as the LSP.
Since $m_{16}$ is so low, Yukawa coupling unification only occurs at the $R\sim 1.09$ level. However,
$\tz_1\tz_1$ can now annihilate through the light Higgs $h$ resonance at a sufficient rate to 
obtain the desired relic density with $\tz_1$ remaining as the LSP\cite{bkss}.

In this paper, we explore the consequences of Yukawa-unified SUSY models for sparticle detection
at the LHC. We focus most of our attention on two cases presented in Table 1 of Ref. \cite{bkss}:
1.~point~A with $m_{16}\sim 9$ GeV and an axino LSP, and 2.~point~D with $m_{16}\sim 3$ TeV and a
neutralino LSP. While our studies here focus on just two cases, we maintain that 
the qualitative features of {\it all} Yukawa-unified SUSY LHC signatures should be 
rather similar to these two cases. In fact, the collider phenomenology of these 
cases is rather similar between the two, since in the first case the neutralino decays 
to an axino far beyond the detector boundaries. Thus, in both cases the lightest neutralino $\tz_1$ 
leads to missing $E_T$ at collider experiments.

For the benefit of the reader, we present in Table \ref{tab:bm} the two case studies we examine. 
We present the parameter space values, sparticle mass spectrum, and in addition the total 
tree-level LHC sparticle production cross section. We also list as percentages some contributing 
$2\to 2$ subprocess reactions. 
%
\begin{table}\centering
\begin{tabular}{lcc}
\hline
parameter & Pt. A & Pt. D \\
\hline
$m_{16}$ & 9202.9 & 2976.5 \\
$m_{1/2}$ & 62.5 & 107.0 \\
$A_0$ & $-19964.5$ & $-6060.3$ \\
$m_{10}$ & 10966.1 & 3787.9 \\
$\tan\beta$ & 49.1 & 49.05 \\
$M_D$ & 3504.4 & 1020.8 \\
$f_t$ & 0.51 & 0.48 \\
$f_b$ & 0.51 & 0.47 \\
$f_\tau$ & 0.52 & 0.52 \\
$\mu$ & 4179.8 & 331.0 \\
$m_{\tg}$   & 395.6 & 387.7 \\
$m_{\tu_L}$ & 9185.4 & 2970.8 \\
$m_{\tst_1}$& 2315.1 & 434.5  \\
$m_{\tb_1}$ & 2723.1 & 849.3 \\
$m_{\te_L}$ & 9131.9 & 2955.8 \\
$m_{\twpm_1}$ & 128.8 & 105.7 \\
$m_{\tz_2}$ & 128.6 & 105.1 \\ 
$m_{\tz_1}$ & 55.6 &  52.6 \\ 
$m_A$       & 3273.6 &  776.8 \\
$m_h$       & 125.4 &  111.1 \\ \hline
$\sigma\ [{\rm fb}]$ & 75579.1 & 89666.1 \\
$\%\ (\tg\tg )$ & 86.8 & 80.5 \\
$\%\ (\twpm_1\tz_2 )$ & 8.8 & 12.8 \\
$\%\ (\tst_1\bar{\tst}_1 )$ & 0 & 1.1 \\
\hline
\end{tabular}
\caption{Masses and parameters in~GeV units
for two cases studies points A and D of Ref. \cite{bkss} 
using Isajet 7.75 with $m_t=171.0$ GeV. We also list the 
total tree level sparticle production cross section 
in fb at the LHC, plus the percent for several
two-body final states.
}
\label{tab:bm}
\end{table}

The remainder of this paper is organized as follows.
In Sec. \ref{sec:csbf}, we present some details involving total sparticle production rates at the LHC 
and also sparticle branching fractions expected in Yukawa-unified SUSY models. 
In Sec. \ref{sec:evgen}, we generate simulated LHC collider events 
associated with these two scenarios along with SM background (BG) rates. 
Since the collider signals are quite similar for both of points A and D, we mainly present figures
for just point A.
Our main findings include:
the Yukawa-unified SUSY scenarios should be easily discoverable at LHC with about 1 fb$^{-1}$ of 
integrated luminosity, {\it without} using missing $E_T$ cuts, by requiring events with a high isolated
lepton multiplicity: $n_\ell\ge 2$ or $3$. 
In Sec. \ref{sec:mass}, we present several distributions which can be used to gain information on sparticle
masses, especially $m_{\tg}$, $m_{\tz_2}$ and $m_{\tz_1}$. The gluino mass $m_{\tg}$ can be extracted from total
cross section results. 
In Sec. \ref{sec:3l}, we discuss a corroborating signal in the $pp\to \twpm_1\tz_2\to 3\ell +\eslt$ channel. 
This channel
comes from the {\it soft} component of signal events involving direct production of the lighter charginos and
neutralinos. 
Our conclusions are presented in Sec.~\ref{sec:conclude}.

\section{Cross sections and branching fractions for sparticles in Yukawa-unified models}
\label{sec:csbf}

Given the characteristic spectrum of superpartners obtained in Yukawa-unified SUSY models,
it is useful to examine what sort of new physics signals we would expect at the LHC.
Obviously, first/second generation squarks and sleptons in the multi-TeV mass range 
will essentially decouple from LHC physics. Gluinos-- in the 350-500 GeV range--
will be produced in abundance via $q\bar{q}$ and $gg$ fusion subprocesses.
Charginos and neutralinos, being in the 100--160 GeV range, may also be produced
with observable cross sections. 

As noted above, we list the tree-level total sparticle production cross sections
obtained from Isajet for cases A and D in Table \ref{tab:bm}. 
In case A, we find $\sigma (tot)\sim 8\times 10^4$ fb,
so that 8000 sparticle pair events are expected at LHC with just 0.1 fb$^{-1}$ of integrated luminosity.
Of this total, 86.7\% comes from gluino pair production, while 8.8\% comes from $\tw_1\tz_2$ 
production and 4.5\% comes from $\tw_1^+\tw_1^-$ production. In case D, $\sigma (tot)\sim 9.6\times 10^4$ fb,
with 80.4\% from $\tg\tg$ production, 12.8\% from $\tw_1\tz_2$ production, $6.4\%$ 
from $\tw_1^+\tw_1^-$ production
while top-squark pair production yields just 1.1\% of the total. Given these production cross sections, we expect
Yukawa-unified  SUSY to yield primarily $\tg\tg$ events at the LHC. Gluino pair production typically 
leads to events with {\it hard} jets, hard $\eslt$ and isolated leptons from the gluino cascade decays\cite{cascade}. 
We also expect a soft component
coming from $\tw_1^+\tw_1^-$ and $\twpm_1\tz_2$ production. While both these reactions lead to
events with rather soft jets, leptons and $\eslt$, 
the latter reaction can also yield clean trilepton events\cite{bcpt_3l},
which should be visible at LHC above SM backgrounds.

For the case of gluino masses other than those listed in Table \ref{tab:bm}, we show in Fig.~\ref{fig:sig_gg}
the total gluino pair production rate versus $m_{\tg}$ at the LHC at tree level (solid) 
and next-to-leading-order (NLO) using
the Prospino program\cite{prospino}. The scale choice is taken to be $Q=m_{\tg}$. 
We take $m_{\tq}$ to be 3 TeV (blue) and 9 TeV (red).
As can be seen, the results hardly vary between this range of squark masses.
The tree level results agree
well with Isajet, but the NLO results typically show an enhancement by a factor $\sim 1.6$.
Thus, we expect Yukawa-unified SUSY models to yield $pp\to\tg\tg X$ events at a 30-150 pb
level at LHC.

In Fig.~\ref{fig:ino}, we show the total -ino pair production cross sections versus
chargino mass $m_{\twpm_1}$. While $\tw_1^\pm\tz_2$ and $\tw_1^+\tw_1^-$ production
dominate, and have rates around $10^3-10^4$ fb over the range of interest, there exists
a sub-dominant rate for $\tw_1^\pm\tz_1$ and also $\tz_1\tz_2$ production.

Now that we see that Yukawa-unified SUSY will yield dominantly gluino pair
production events at the LHC, we next turn to the gluino branching fractions
in order to understand their event signatures. All sparticle branching
fractions are calculated with Isajet 7.75. In Fig.~\ref{fig:bf}, we show 
various gluino branching fractions for points A and D.
We see immediately that in both cases, $BF(\tg\to b\bar{b}\tz_2)$ dominates
at around 56\%. This is followed by $BF(\tg\to b\bar{b}\tz_1)$ at $\sim 16\%$,
and $BF(\tg\to b\bar{t}\tw_1^+)$ and $BF(\tg\to t\bar{b}\tw_1^-)$ each at
$\sim 10\%$. Decays to first and second generation quarks are much suppressed due to the 
large first and second generation squark masses. 
From these results, we expect gluino pair production events to be
rich in $b$-jets, $\eslt$ and occassional isolated leptons from the leptonic
decays $\tw_1\to \ell\bar{\nu}_\ell\tz_1$ and $\tz_2\to \ell^+\ell^-\tz_1$, where
$\ell =e$ or $\mu$.

\FIGURE[htb]{
\epsfig{file=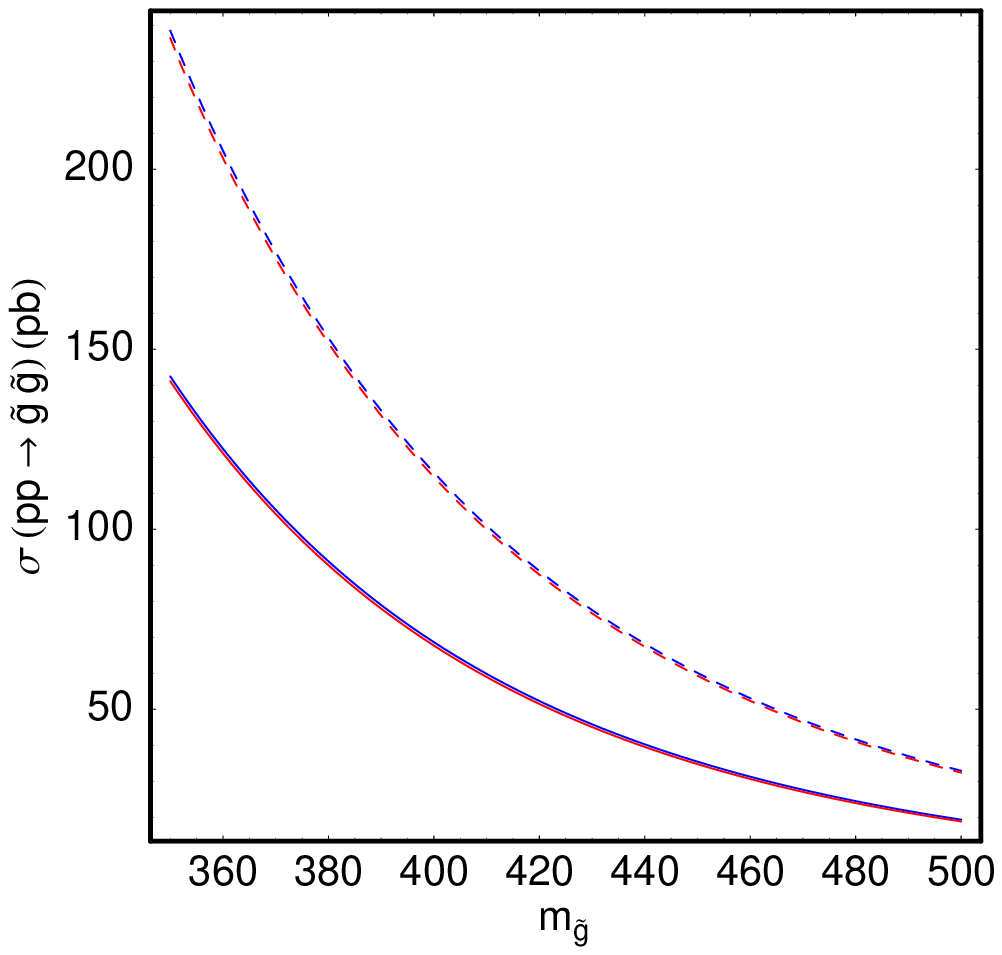,width=10cm} 
\caption{ Plot of $\sigma (pp\to \tg\tg X)$ in pb at 
$\sqrt{s}=14$ TeV versus $m_{\tg}$. We use Prospino with 
scale choice $Q=m_{\tg}$, and show LO (solid) and NLO (dashes)
predictions in the vicinity of point A (red) and point D (blue)
from Table \ref{tab:bm}.
}\label{fig:sig_gg}}

\FIGURE[htb]{
\epsfig{file=ino.eps,width=10cm} 
\caption{ Plot of various -ino pair production 
processes in fb at $\sqrt{s}=14$ TeV versus $m_{\twpm_1}$,
for $m_{\tq}=3$ TeV and $\mu =m_{\tg}$, with $\tan\beta =49$
and $\mu >0$.}
\label{fig:ino}}

\clearpage

\FIGURE[t]{
\epsfig{file=bf-ptAptD.eps,width=14cm} 
\caption{ Plot of various sparticle branching fractions
taken from Isajet for points A and D from Table \ref{tab:bm}.
}\label{fig:bf}}

\section{Gluino pair production signals at the LHC}
\label{sec:evgen}

To examine collider signals from Yukawa-unified SUSY at the LHC in more detail, 
we generate $10^6$ sparticle pair production events for points A and D, 
corresponding to $13$ and $11$ fb$^{-1}$ of integrated luminosities.
We use Isajet 7.75\cite{isajet} for the simulation of signal and 
background events at the LHC. A toy detector simulation is employed with
calorimeter cell size
$\Delta\eta\times\Delta\phi=0.05\times 0.05$ and $-5<\eta<5$. The HCAL
energy resolution is taken to be $80\%/\sqrt{E}+3\%$ for $|\eta|<2.6$ and
FCAL is $100\%/\sqrt{E}+5\%$ for $|\eta|>2.6$. 
The ECAL energy resolution
is assumed to be $3\%/\sqrt{E}+0.5\%$. We use a UA1-like jet finding algorithm
with jet cone size $R=0.4$ and require that $E_T(jet)>50$ GeV and
$|\eta (jet)|<3.0$. Leptons are considered
isolated if they have $p_T(e\ or\ \mu)>20$ GeV and $|\eta|<2.5$ with 
visible activity within a cone of $\Delta R<0.2$ of
$\Sigma E_T^{cells}<5$ GeV. The strict isolation criterion helps reduce
multi-lepton backgrounds from heavy quark ($c\bar c$ and $b\bar{b}$) production.
We also invoke a lepton identification efficiency of 75\% for leptons with 
20 GeV$<p_T(\ell )<$ 50 GeV, and 85\% for leptons with $p_T(\ell )>50$ GeV.

We identify a hadronic cluster with $E_T>50$ GeV and $|\eta(j)|<1.5$
as a $b$-jet if it contains a $B$ hadron with $p_T(B)>15$ GeV and
$|\eta (B)|<3$ within a cone of $\Delta R<0.5$ about the jet axis. We
adopt a $b$-jet tagging efficiency of 60\%, and assume that
light quark and gluon jets can be mis-tagged as $b$-jets with a
probability $1/150$ for $E_T\le 100$ GeV, $1/50$ for $E_T\ge 250$ GeV, 
with a linear interpolation for $100$ GeV$<E_T<$ 250 GeV\cite{xt}. 

In addition to signal, we have generated background events using Isajet for
QCD jet production (jet-types include $g$, $u$, $d$, $s$, $c$ and $b$
quarks) over five $p_T$ ranges as shown in Table \ref{tab:bg}. 
Additional jets are generated via parton showering from the initial and final state
hard scattering subprocesses.
We have also generated backgrounds in the $W+jets$, $Z+jets$, 
$t\bar{t}$ (with $m_t=171$ GeV) and $WW,\ WZ,\ ZZ$ channels at the rates shown in 
Table \ref{tab:bg}. The $W+jets$ and $Z+jets$ backgrounds
use exact matrix elements for one parton emission, but rely on the 
parton shower for subsequent emissions.
%
\begin{table}
\begin{center}
\begin{tabular}{lrcccc}
\hline
\multicolumn{2}{c}{process} & events & $\sigma$ (fb) & $\c1p$ & $\c1p +\eslt$  \\
\hline
\multicolumn{2}{l}{QCD ($p_T:0.05-0.1$ TeV)} & $10^6$ & $2.6\times 10^{10}$ & $4.1\times 10^5$ & --  \\
\multicolumn{2}{l}{QCD ($p_T:0.1-0.2$ TeV)} & $10^6$ & $1.5\times 10^{9}$ & $1.4\times 10^7$ & --  \\
\multicolumn{2}{l}{QCD ($p_T:0.2-0.4$ TeV)} & $10^6$ & $7.3\times 10^{7}$ & $6.5\times 10^6$ & 2199  \\
\multicolumn{2}{l}{QCD ($p_T:0.4-1.0$ TeV)} & $10^6$ & $2.7\times 10^{6}$ & $2.8\times 10^5$ & 1157 \\
\multicolumn{2}{l}{QCD ($p_T:1-2.4$ TeV)} & $10^6$ & $1.5\times 10^{4}$ & $1082$ & 25  \\
\multicolumn{2}{l}{${W\to \ell\nu_\ell +jets}$} & $5\times 10^5$ & $3.9\times 10^{5}$ & $ 3850$ & 1275  \\
\multicolumn{2}{l}{${Z\to\tau\bar{\tau}+jets}$} & $5\times 10^5$ & $1.4\times 10^{5}$ & $1358$ & 652  \\
\multicolumn{2}{l}{$t\bar{t}$} & $3\times 10^6$ & $4.9\times 10^{5}$ & $8.2\times 10^4$ & 2873  \\
\multicolumn{2}{l}{$WW,ZZ,WZ$} & $5\times 10^5$ & $8.0\times 10^{4}$ & $197$ & 7 \\
\hline
\multicolumn{2}{l}{Total BG} & $9.5\times 10^6$ & $2.76\times 10^{10}$ & $2.13\times 10^{7}$ & 8188  \\ 
\hline
Point A: &  & $10^6$ & $7.6\times 10^{4}$ & $3.6\times 10^4$ & 8914 \\
 & $S/B\rightarrow$ & -- & -- & 0.002 & 1.09  \\
 & $S/\sqrt{S+B}$ $(1\,\rm fb^{-1})\rightarrow$ & -- & -- & -- & 68  \\
Point D: & & $10^6$ & $9.0\times 10^{4}$ & $3.7\times 10^4$ & 10843  \\
 & $S/B\rightarrow$ & -- & -- & 0.002 & 1.32  \\
 & $S/\sqrt{S+B}$ $(1\,\rm fb^{-1})\rightarrow$ & -- & -- & -- & 78  \\
\hline
\end{tabular}
\caption{Events generated and cross sections (in fb) for various SM background 
and signal processes before and after cuts.
The $\c1p$ and $\eslt$ cuts are specified in the text. The $W+jets$
and $Z+jets$ background has been computed within the restriction
$p_T(W,Z)>100$ GeV.}
\label{tab:bg}
\end{center}
\end{table}

First we require modest cuts: $n(jets)\ge 4$. Also, SUSY events are 
expected to spray large $E_T$ throughout the calorimeter, while QCD dijet events 
are expected to be typically back-to-back. Thus, we expect QCD background to 
be peaked at transverse sphericity $S_T\sim 0$, while SUSY events have larger values of $S_T$. 
\footnote{Here, $S_T$ is the usual sphericity variable, restricted to the transverse plane, as is appropriate
for hadron colliders.  Sphericity matrix is given as
\begin{equation}
S = \left( \begin{array}{cc}
\sum p_x^2 & \sum p_x p_y \\
\sum p_x p_y & \sum p_y^2 \\
\end{array} \right)
\end{equation}
from which $S_T$ is defined as $2\lambda_1 / (\lambda_1 + \lambda_2)$, where 
$\lambda_{1,2}$ are the larger and smaller eigenvalues of $S$.} 
The actual $S_T$ distribution for point A is shown in Fig.~\ref{fig:st}   
(the $S_T$ distribution for point D is practically the same).
Motivated by this, 
we require $S_T>0.2$ to reject QCD-like events.

We plot the jet $E_T$ distributions of the four highest $E_T$ jets from Pt. A (in color) and the total 
SM background (gray histogram) in Fig.~\ref{fig:etjets}, ordered from highest to lowest $E_T$, with jets 
labelled as $j1-j4$.  The histograms are normalized to unity in order to clearly see the differences in 
distribution shapes. Again, the distributions for point D look very similar. We find that
the highest $E_T$ jet distribution peaks around $E_T\sim 150$ GeV with a 
long tail extending to higher $E_T$ values, while for the background it peaks at a lower value of 
$E_T\sim 100$ GeV. Jet 2 and jet 3 have peak distributions around 100 GeV both for the signal and 
backgrounds, while the jet 4 distribution backs up against the minimum jet $E_T$ requirement that 
$E_T(jet)>50$ GeV. Thus, at little cost to signal but with large BG rejection, we require 
$E_T(j1)>100$ GeV.

The collection of cuts so far is dubbed $\c1p$\cite{bps}:
\\
\\
\textbf{C1$^\prime$ cuts:}
\bea
n(jets) &\ge & 4,\label{eq:c1} \\
E_T(j1,j2,j3,j4)& \ge & 100,\ 50,50,50\ {\rm GeV},\\
S_T &\ge &0.2 .
\label{eq:c3}
\eea

\FIGURE[htb]{
\epsfig{file=so10ptA_C1p-tsph.eps,width=10cm} 
\caption{Plot of distribution in transverse sphericity $S_T$ for events
with cuts $\c1p$ from benchmark point A and the summed SM background; 
point D leads to practically the same distribution. 
}\label{fig:st}}

\FIGURE[htb]{
\epsfig{file=so10ptA_C1p-etjets.eps,width=14cm} 
\caption{Plot of jet $E_T$ distributions for events
with $\ge 4$ jets after requiring just $S_T>0.2$, from 
benchmark point A; distributions for point D are nearly the same.
}\label{fig:etjets}}

The classic signature for SUSY collider events is the presence of
jets plus large $\eslt$\cite{etmiss}. In Fig.~\ref{fig:etmiss}, we
show the expected distribution of $\eslt$ from points A and D, along with SM
BG. We do see that signal becomes comparable to BG 
around $\eslt\sim 150$ GeV. We list cross sections  from the two signal cases plus SM backgrounds
in Table \ref{tab:bg} after cuts $\c1p$ plus $\eslt >150$ GeV. While signal $S$ is 
somewhat higher than the summed BG $B$, the signal and BG rates are rather comparable in this case:
$S/B=1.09$ for pt.\ A while $S/B=1.32$ for pt.\ D.

Even so, it has been noted in Ref. \cite{bps}
that $\eslt$ may be a difficult variable to reliably construct during the
early stages of LHC running. The reason is that 
missing transverse energy can arise not only from the presence of 
weakly interacting neutral particles such as neutrinos or the lightest
neutralinos, but also from a variety of other sources, including:
\bi
\item energy loss from cracks and un-instrumented regions of the detector,
\item energy loss from dead cells,
\item hot cells in the calorimeter that report an energy deposition
even if there is not one,
\item mis-measurement in the electromagnetic
calorimeters, hadronic calorimeters or muon detectors,
\item real missing transverse energy produced in jets due to semi-leptonic decays of heavy flavors,
\item muons and 
\item the presence of mis-identified cosmic rays in events.
\ei
Thus, in order to have a solid grasp of expected $\eslt$ from
SM background processes, it will be necessary to have detailed knowledge of the
complete detector performance. As experience from the Tevatron suggests,
this complicated task may well take some time to complete. 
The same may also be true at the LHC, 
as many SM processes will have to be scrutinized first 
in order to properly calibrate the detector\cite{calibrate}. 
For this reason, SUSY searches using $\eslt$
as a crucial requirement may well take rather longer than a 
year to provide reliable results.
 
For this reason, Ref. \cite{bps} advocated to look for 
SUSY signal events by searching for a high multiplicity
of detected objects, rather than inferred undetected objects, 
such as $\eslt$.
In this vein, we show in Fig.~\ref{fig:njets} the jet multiplicity 
from SUSY signal (Pts.\ A and D) along with SM BG after cuts $\c1p$,
{\it i.e} with {\it no} $\eslt$ cut. 
We see that at low jet multiplicity, SM BG dominates the SUSY 
signal. However, signal/background increases with
$n(jets)$ until at $n(jets)\sim 15$ finally signal overtakes BG in 
raw rate.

One can do better in detected $b$-jet multiplicity, $n_b$, 
as shown in Fig.~\ref{fig:nbjets}. Since each gluino is expected to decay
to two $b$-jets, we expect a high $n_b$ multiplicity in signal.
In this case, BG dominates signal at low $n_b$, but signal overtakes 
BG around $n_b\simeq 4$.

The isolated lepton multiplicity $n_\ell$ is shown in Fig.~\ref{fig:nleps}
for signal and SM BG after cuts $\c1p$. In this case, isolated leptons
should be relatively common in gluino cascade decays. 
We see that signal exceeds BG already at $n_\ell =2$, and far exceeds BG at 
$n_\ell =3$:
at $n_\ell =3$, we have 35 (42) fb of signal cross section after cuts for point A (D) and no BG.
In fact, high isolated lepton multiplicity was advocated
in Ref. \cite{bps} in lieu of an $\eslt$ cut to search for 
SUSY with integrated luminosities of around $1~{\rm fb}^{-1}$ at LHC.

We also point out here that $\tg\tg$ production can lead to large rates for
same-sign (SS) isolated dilepton production\cite{ssdilep}, while SM BG for this
topology is expected to be small.
We plot in Fig.~\ref{fig:ss} the rate of events from signal and 
SM BG for cuts $\c1p$ plus a pair of isolated SS dileptons, versus
jet multiplicity. While BG is large at low $n(jets)$, signal emerges 
from and dominates BG at higher jet multiplicities.

\FIGURE[htb]{
\epsfig{file=so10ptsAD_C1prime-etmiss.eps,width=10cm} 
\caption{Plot of missing $E_T$ for events with 
$\ge 4$ jets after cuts $\c1p$, from benchmark points A (full red line) 
and D (dashed blue line).
}\label{fig:etmiss}}

\FIGURE[htb]{
\epsfig{file=so10ptsAD_C1prime-njets.eps,width=10cm} 
\caption{Plot of jet multiplicity from benchmark points A (full red line) 
and D (dashed blue line) after cuts $\c1p$ along with SM backgrounds.
}\label{fig:njets}}

\FIGURE[htb]{
\epsfig{file=so10ptsAD_C1prime-nbjets.eps,width=10cm} 
\caption{Plot of $b$-jet multiplicity from benchmark points A (full red line) 
and D (dashed blue line) after cuts $\c1p$ along with SM backgrounds.
}\label{fig:nbjets}}

\FIGURE[htb]{  
\epsfig{file=so10ptsAD_C1prime-nleps.eps,width=10cm} 
\caption{Plot of isolated lepton multiplicity from benchmark points A (full red line) 
and D (dashed blue line) after cuts $\c1p$ along with SM backgrounds.
}\label{fig:nleps}}

\FIGURE[htb]{
\epsfig{file=so10ptsAD-SS-njets.eps,width=10cm} 
\caption{Plot of jet multiplicity in events with
isolated SS dileptons from benchmark points A (full red line) 
and D (dashed blue line) after cut $S_T>0.2$ along with SM backgrounds.
}\label{fig:ss}}

\clearpage 

\section{Sparticle masses from gluino pair production}
\label{sec:mass}

There exists good prospects for sparticle mass measurements 
in Yukawa-unified SUSY models at the LHC. One reason is that
sparticle pair production is dominated by a single reaction:
gluino pair production. The other propitious circumstance is
that the mass difference $m_{\tz_2}-m_{\tz_1}$ is highly favored 
to be bounded by $M_Z$. This means that $\tz_2$ decays dominantly into
three body modes such as $\tz_2\to\tz_1\ell\bar{\ell}$ with
a significant branching fraction, while the so-called ``spoiler decay modes''
$\tz_2\to\tz_1 Z$ and $\tz_2\to \tz_1 h$ are kinematically closed. 
The three body decay mode is important in that it yields
a continuous distribution in $m(\ell\bar{\ell})$ which
is bounded by $m_{\tz_2}-m_{\tz_1}$: this kinematic edge can serve as the starting point
for sparticle mass reconstruction in cascade decay events\cite{mlledge,bcpt_3l}.

As an example, we require cuts set $\c1p$ plus the presence of a pair
of same-flavor opposite-sign (SF/OS) isolated leptons.  
The resulting dilepton invariant mass distributions for points A and D are shown
in Fig.~\ref{fig:mllb}.  Furthermore, in Fig.~\ref{fig:mllbsub}, we plot the different-flavor subtracted 
distributions: $d\sigma /dm(\ell^+\ell^- )-d\sigma /dm(\ell^+\ell^{-\prime})$, which allow for a 
subtraction of $e^+\mu^-$ and $e^-\mu^+$ pairs from processes like chargino pair production in cascade 
decay events. A clear peak at $m(\ell\bar{\ell})=M_Z$ is seen in the BG distribution. This comes mainly 
from QCD jet production events, since Isajet includes $W$ and $Z$ radiation in its 
parton shower algorithm (in the effective $W$ approximation). The signal displays a histogram easily 
visible above SM BG with a distinct cut-off at $m_{\tz_2}-m_{\tz_1}=73$ (52.5) GeV for point A (D). 
Isajet contains
the exact decay matrix elements in 3-body decay processes, and in these cases
we see distributions that differ from pure phase space, and yield distributions skewed to higher 
$m(\ell\bar{\ell})$ values. This actually shows the influence of the virtual $Z$ in the decay diagrams,
since the decay distribution is dominated by $Z^*$ exchange. The 
closer $m_{\tz_2}-m_{\tz_1}$ gets to $M_Z$, the more the $Z$-boson
propagator pulls the dilepton mass distribution towards $M_Z$\cite{new3l}.
The dilepton mass edge should be measureable to a precision of $\sim 50$ MeV according to  
Ref. \cite{edgefit}.

\FIGURE[p!]{
\epsfig{file=so10ptsAD_C1p-SFOS.eps,width=11cm}
\caption{SF/OS dilepton invariant mass distribution after cuts $\c1p$ 
from benchmark points A (full red line) and D (dashed blue line)
along with SM backgrounds. \vspace*{1cm}
}\label{fig:mllb}}

\FIGURE[p!]{
\epsfig{file=so10ptsAD-OSdiff.eps,width=11cm} 
\caption{Same as Fig.~\ref{fig:mllb} but for same-flavor minus
different-flavor subtracted invariant-mass.
}\label{fig:mllbsub}}

For Yukawa-unified SUSY models, the branching fraction
$BF(\tg\to b\bar{b}\tz_2)$ dominates at around 56\%. 
If one can identify events with a clean $\tz_2\to\ell\bar{\ell}\tz_1$
decay, then one might also try to extract the invariant mass of 
the associated two $b$-jets coming from the gluino decay, 
which should have a kinematic upper edge at $m_{\tg}-m_{\tz_2}\simeq 267$ 
($283$) GeV for point A (D).  
A second, less pronounced endpoint is expected at  $m_{\tg}-m_{\tz_1}\simeq 340$ 
($335$) GeV due to $\tg\to b\bar{b}\tz_1$ decays which have $\sim 16\%$ branching ratio.
A third endpoint can also occur from $\tz_2\to\tz_1b\bar{b}$ decay where
$m_{\tz_2}-m_{\tz_1}=73$ (52.5) GeV, respectively.

The high multiplicity of $b$-jets (typically two from each gluino decay) however  
poses a serious combinatorics problem in extracting the $b\bar b$ invariant-mass 
distribution. 
In a first attempt, we required at least two tagged $b$-jets   
along with cuts $\c1p$ and SF/OS dileptons,   
and plotted the {\it minimum} invariant mass of all the $b$-jets in the 
event. 
The resulting distributions for points A and D peaked 
around $m(b\bar{b})\sim 100$ GeV, with a distribution tail extending well beyond
the above-mentioned kinematic endpoints. It was clear that we were frequently 
picking up wrong $b$-jet pairs, with either each $b$ originating from a different gluino, 
or one or both $b$'s originating from a $\tz_2\to\tz_1 b\bar{b}$ decay.
A more sophisticated procedure to pair the correct $b$-jets is required 
for the $b\bar b$ invariant-mass distribution. 

\clearpage
Parton-level Monte Carlo simulations revealed that the two hardest (highest $E_T$)
$b$-jets almost always originated from {\it different} gluinos. Thus, 
we require events with at least four tagged $b$-jets (along with cuts $\c1p$)
and combine the hardest $b$-jet with either the 3rd or 4th hardest $b$-jet,
creating an object $X_1(bb)$. Moreover, we combine the 2nd hardest $b$-jet with
the 4th or 3rd hardest $b$-jet, creating an object $X_2(bb)$. Next, we calculate 
$\Delta m(X_1-X_2)\equiv |m(X_1) -m(X_2)|/(m(X_1)+m(X_2))$. 
We select the set of $bb$ clusters which has the 
minimum value of $\Delta m(X_1-X_2)$, and plot the invariant mass of both the clusters.
This procedure produces a sharp kinematic edge in $m(bb)$ in parton level 
simulations.\footnote{We have also tried other methods such as picking 
the $X_1 - X_2$ pair with maximum $\Delta \phi(X_1 - X_2)$, maximum $\Delta R(X_1 - X_2)$, 
minimum $\delta p_T(X_1 - X_2)$ and 
minimum average invariant mass $avg(m(X_1), m(X_2))$.  We also tried to separate the 
resulting $b$ jets into hemispheres.  In the end, the best amount of correct assignment was 
achieved with the choice of the $\Delta m(X_1-X_2)_{min}$.}
The resulting distribution from Isajet is shown in Fig.~\ref{fig:mbbXY} for points A and D. 
The distribution peaks at a higher value of $m(X_i)$ ($i=1,2$), and is largely
bounded by the kinematic endpoints, although a tail still extends to high 
$m(X_i)$. Part of the high $m(X_i)$ tail is due to the presence of 
$\tg\to\tz_1 b\bar{b}$ decays, which have a higher kinematic endpoint than the
$\tg\to \tz_2 b\bar{b}$ decays, and of $\tz_2\to\tz_1 b\bar{b}$ decays. 
In addition, there is a non-negligible background contribution, 
indicated by the gray histogram. 

We can do much better, albeit with reduced statistics, by requiring in addition 
the presence of a pair of SF/OS dileptons.  
Applying the same procedure as described above, we
arrive at the distribution shown in Fig.~\ref{fig:mbbXYSFOS}.
In this case, the SM BG is greatly reduced, and two mass edges begin to appear.

\FIGURE[ht!]{
\epsfig{file=so10ptsAD_C1p-ge4b-mbbdelXYmin.eps,width=11cm} 
\caption{Plot of $m(X_{1,2})$  from benchmark points A and D
along with SM backgrounds in events with cuts $\c1p$ plus
$\ge 4$ $b$-jets and minimizing $\Delta m(X_1-X_2)$; see text for details.
}\label{fig:mbbXY}}

\FIGURE[ht!]{
\epsfig{file=so10ptsAD_C1p-SFOSge4b-mbbdelXYmin.eps,width=11cm} 
\caption{Same as Fig.~\ref{fig:mbbXY} but requiring in addition a pair of SF/OS leptons.
}\label{fig:mbbXYSFOS}}

It should also be possible to combine the invariant mass
of the SF/OS dilepton pair with a $b\bar{b}$ pair. 
Requiring cuts $\c1p$ plus $\ge 2$ $b$-jets 
plus a pair of SF/OS dileptons (with $m(\ell\bar{\ell})<m_{\tz_2}-m_{\tz_1}$), 
we reconstruct $m(bb\ell\bar{\ell})$.
The result is shown in Fig.~\ref{fig:mbbll}. While the distribution 
peaks at $m(\ell\bar{\ell}b\bar{b})\sim 300$ GeV, a kinematic edge
at $m_{\tg}-m_{\tz_1}\sim 340$ GeV is also visible (along with 
a mis-identification tail extending to higher invarant masses). 

We try to do better, again with a loss of statistics, by
requiring $\ge 4$ $b$-jets instead of $\ge 2$, and combining the $bb$
clusters into objects $X_1$ and $X_2$ as described above.
We again choose the set of clusters which give the minimum of
$\Delta m(X_1-X_2)$ and combine each of these clusters with the
$\ell\bar{\ell}$ pair. We take the minimum of the two
$m(X_i \ell\bar{\ell})$ values, and plot the distribution in Fig.~\ref{fig:mbbllXY}. 
In this case, the SM BG is even more reduced, and
the $m_{\tg}-m_{\tz_1}$ mass edge seems somewhat more apparent.

According to \cite{edgefit}, measurements of hadronic mass edges can be made with 
a precision of roughly 10\%. Nevertheless, from the kinematic distributions discussed
above we can only determine mass differences.
There is still not enough information to extract absolute masses, i.e.\ each of
$m_{\tg}$, $m_{\tz_2}$ and $m_{\tz_1}$. 
However, it is pointed out in Ref. \cite{gabe} that in cases 
(such as the focus point region of minimal supergravity) where sparticle pair
production occurs nearly purely from $\tg\tg$ production, and when the dominant
$\tg$ branching fractions are known (from a combination of theory and experiment), 
then the total $\tg\tg$ production cross section after cuts allows
for an absolute measurement of $m_{\tg}$ to about an 8\% accuracy. These conditions should apply
to our Yukawa-unified SUSY cases, if we assume the $\sim 56\%$ branching fraction for
$\tg\to b\bar{b}\tz_2$ decay (from theory). The study of Ref. \cite{gabe} required that one fulfill the cuts
$C2$ which gave robust gluino pair production signal along with small SM backgrounds:

\FIGURE[p!]{
\epsfig{file=so10ptsAD_C1p-SFOS-mllbbmin.eps,width=11cm} 
\caption{Plot of $m(bb\ell^+\ell^- )_{min}$  from points A and D 
along with SM backgrounds.
}\label{fig:mbbll}}
\vspace*{1cm}

\FIGURE[p!]{
\epsfig{file=so10ptsAD_C1p-SFOSge4b-mbblldelXYmin.eps,width=11cm} 
\caption{Plot of $m(X_{1,2}\ell^+\ell^- )_{min}$ from points A and D, 
minimizing $\Delta m(X_1-X_2)$ as explained in the text, 
along with SM backgrounds.
}\label{fig:mbbllXY}}

\clearpage

\noindent
\textbf{C2 cuts:}
\bea
\eslt > (max(100\ {\rm GeV}, 0.2 M_{eff}), \\
n(jets)\ge 7,\\
n(b-jets)\ge 2,\\
E_T(j1,j2-j7)>100,\ 50\ {\rm GeV},\\
A_T>1400\  {\rm GeV},\\
S_T \ge 0.2\  ,
\label{eq:c2}
\eea
%
where $A_T$ is the augmented effective mass $A_T=\eslt +\sum_{leptons}E_T +\sum_{jets} E_T$.
In this case, the summed SM background was about 1.6 fb, while signal rate for Point A (D) is
57.3 (66.2) fb.
The total cross section after cuts varies strongly with $m_{\tg}$, allowing an extraction of $m_{\tg}$
to about 8\% for 100 fb$^{-1}$ integrated luminosity, after factoring in QCD and branching fraction
uncertainties in the total rate. Once an absolute value of $m_{\tg}$ is known, then $m_{\tz_2}$
and $m_{\tz_1}$ can be extracted to about 10\% accuracy from the invariant mass edge information.

\section{Trilepton signal from $\tw_1\tz_2$ production}
\label{sec:3l}

While the signal from gluino pair production at the LHC from 
Yukawa-unified SUSY models will be 
very robust, it will be useful to have a confirming SUSY signal in an 
alternative channel. From Fig.~\ref{fig:ino}, we see that there also exists
substantial cross sections for $\tw_1^\pm\tz_1$, $\tw_1^+\tw_1^-$ and $\tw_1^\pm\tz_2$
production. The $\twpm_1\to\tz_1 f\bar{f}'$ and $\tz_2\to\tz_1 f\bar{f}$ decays
(here $f$ stands for any of the SM fermions) are dominated by $W$ and $Z$ exchange,
respectively, so that in this case the branching fractions  $BF(\twpm_1\to\tz_1 f\bar{f}')$
are similar to $BF(W^\pm\to f\bar{f}')$ and $BF(\tz_2\to\tz_1 f\bar{f})$ is similar to
$Z\to f\bar{f}$. 

The $\tw_1^\pm\tz_1\to\tz_1 q\bar{q}'+\tz_1$ process will be difficult to observe at LHC
since the final state jets and $\eslt$ will be relatively soft, and likely buried under SM background.
Likewise, the $\tw_1^-\tz_1\to\tz_1 \ell\bar{\nu}_\ell\tz_1$ signal will be buried under
a huge BG from $W\to\ell\bar{\nu}_\ell$ production. The $\tw_1^+\tw_1^-$ production reaction will also be difficult
to see at LHC. The purely hadronic final state will likely be buried under QCD and $Z+jets$ BG, 
while the lepton plus jets final state will be buried under $W+jets$ BG. The dilepton final state will
be difficult to extract from $W^+W^-$ and $t\bar{t}$ production. 

The remaining reaction, $\tw_1^\pm\tz_2$ production, yields a trilepton final state 
from $\twpm_1\to\tz_1\ell\bar{\nu}_\ell$ and $\tz_2\to\tz_1\ell\bar{\ell}$ decays which in many cases
is observable above SM BG. The LHC reach for $\twpm_1\tz_2\to 3\ell +\eslt$  production was
mapped out in Ref. \cite{bcpt}, and the reach was extended into the 
hyperbolic branch/focus point (HB/FP) region in Ref. \cite{bkpu}. The method was to use the cut set
SC2 from Ref. \cite{new3l} but as applied to the LHC. For the clean trilepton signal from
$\tw_1^\pm\tz_2\to 3\ell +\eslt$ production, we require:
\bi
\item three isolated leptons with $p_T(\ell )>20$ GeV and $|\eta_\ell |<2.5$,
\item SF/OS dilepton mass 20 GeV $<m(\ell^+\ell^- )<81$ GeV, to avoid BG from
photon and $Z$ poles in the $2\to 4$ process $q\bar{q}'\to \ell\bar{\ell}\ell'\bar{\nu}_\ell$ ,
\item a transverse mass veto 60 GeV $<M_T(\ell,\eslt )<85$ GeV to reject on-shell $W$
contributions, 
\item $\eslt >25$ GeV and,
\item veto events with $n(jets)\ge 1$.
\ei 

The resulting BG levels and signal rates for points A and D are listed in Table \ref{tab:3l}.
The $2\to 2$ processes are calculated with Isajet, while the $2\to 4$ processes are
calculated at parton level using Madgraph1\cite{madgraph}.
The combination of hard lepton $p_T$ cuts and the requirement that $n(jets)=0$
leaves us with no $2\to 2$ background, while the parton level $2\to 4$ BG remains at $0.7$ fb.
Here, we see that signal from the two Yukawa-unified points well exceeds background. 

In the clean $3\ell$ channel, since two of the leptons ought to come from $\tz_2\to\tz_1\ell\bar{\ell}$ decay, 
they should display a confirmatory dilepton mass edge at $m_{\tz_2}-m_{\tz_1}$ as is evident
in the gluino pair production events, where the dileptons are accompanied by high jet 
multiplicity.
The distribution in $m(\ell^+\ell^- )$ is shown in Fig.~\ref{fig:mll_3l}.
Event rates are seen to be lower than those from $\tg\tg$ production.  Integrated luminosity needed for 
a discovery with $5\sigma$ significance would be $2.83$ fb$^{-1}$ for pt. A and $1.5$ fb$^{-1}$ for pt. 
D\footnote{Significance is defined as $S/\sqrt{(S+B)}$}. 

\begin{table}
\begin{center}
\begin{tabular}{lrccc}
\hline
\multicolumn{2}{l}{process} & events & $\sigma$ (fb) & after cuts (fb)  \\
\hline
\multicolumn{2}{l}{$t\bar{t}$} & $3\times 10^6$ & $4.9\times 10^{5}$ & -- \\
\multicolumn{2}{l}{$WW,ZZ,WZ$} & $5\times 10^5$ & $8.0\times 10^{4}$ & -- \\
\multicolumn{2}{l}{$W^*Z^*,\ W^*\gamma^*\to \ell\bar{\ell}\ell'\nu_{\ell'}$} & $10^6$ & -- & 0.7 \\
\hline
\multicolumn{2}{l}{Total BG} & $4.5\times 10^5$ & -- & 0.7 \\
\hline
Point\ A: & -- & $10^6$ & $7.6\times 10^{4}$ &  3.4 \\
 & $S/B\rightarrow$ & -- & -- & 4.86 \\
 & $S/\sqrt{S+B}$ $(10\,\rm fb^{-1})\rightarrow$ & -- & -- & 5.31\\
Point\ D: & -- & $10^6$ & $9.0\times 10^{4}$ &  4.1 \\
 & $S/B\rightarrow$ & -- & -- & 5.86 \\
 & $S/\sqrt{S+B}$ $(10\,\rm fb^{-1})\rightarrow$ & -- & -- & 5.92\\
\hline 
\end{tabular}
\caption{Clean trilepton signal after cuts listed in the text.}
\label{tab:3l}
\end{center}
\end{table}
%

\FIGURE[ht]{
\epsfig{file=so10ptsAD_SC2-0j-trilep-mll.eps,width=11cm} 
\caption{Plot of $m(\ell^+\ell^- )$  in the clean trilepton
channel from points A and D along with SM backgrounds.
}\label{fig:mll_3l}}

\section{Summary and conclusions}
\label{sec:conclude}

Simple SUSY grand unified models based on the gauge group $SO(10)$
may have $t-b-\tau$ Yukawa coupling unification in addition to
gauge group and matter unification. By assuming the MSSM is the effective field theory 
valid below $M_{GUT}$, we can, starting with weak scale fermion masses 
as boundary conditions, check whether or not these third generation 
Yukawa couplings actually unify. The calculation depends sensitively on the 
entire SUSY particle mass spectrum, mainly through radiative corrections to the $b$, $t$ and
$\tau$ masses. It was found in previous works that $t-b-\tau$ Yukawa coupling unification can 
occur, but only for very restrictive soft SUSY breaking parameter boundary conditions valid at the
GUT scale, leading to a radiatively induced inverted mass hierarchy amongst the sfermion masses.
While squarks and sleptons are expected to be quite heavy, gluinos, winos and binos are
expected to be quite light, and will be produced at large rates at the CERN LHC. 

We expect LHC collider events from Yukawa-unified SUSY models to be dominated by gluino pair 
production at rates of $(30-150)\times 10^3$ fb. The $\tg$s decay via 3-body modes into
$b\bar{b}\tz_2$, $b\bar{b}\tz_1$ and $tb\twpm_1$, followed by  leptonic or hadronic
3-body decays of the $\tz_2$ and $\twpm_1$. A detailed simulation of signal and SM BG
processes shows that signal should be easily visible above SM BG in the $\ge 4$ jets
plus $\ge 3\ell$ channel, even without using the $\eslt$ variable, with about 1 fb$^{-1}$ of integrated 
luminosity. 

If Yukawa-unified signals from $\tg\tg$ production are present, then 
at higher integrated luminosities, mass edges in the
$m(\ell^+\ell^- )$, $m(b\bar{b})$ and $m(b\bar{b}\ell^+\ell^- )$ channels 
along with total cross section rates (which depend sensitively on the value of
$m_{\tg}$) should allow
for sparticle mass reconstruction of $m_{\tg}$, $m_{\tz_2}$ and $m_{\tz_1}$
to ${\cal O}(10\%)$ accuracy for $\sim 100$ fb$^{-1}$ of integrated luminosity.
The gluino pair production signal can be corroborated by another 
signal in the clean trilepton channel from $\tw_1\tz_2\to 3\ell +\eslt$, which
should also be visible at higher integrated luminosities.
Thus, based on the study presented here, 
we expect LHC to either discover or rule out $t-b-\tau$ Yukawa-unified SUSY models 
within the first year or two of operation.

\acknowledgments

This research was supported in part by the U.S. Department of Energy
grant numbers DE-FG02-97ER41022.  
This work is also part of the French ANR project ToolsDMColl, BLAN07-2-194882.
SS acknowledges financial support by Turkish Atomic Energy Authority.
SK thanks the CERN Theory unit for hospitality during stages of this work.

%

\end{document}